\documentclass[9, journal=jacsat,manuscript=article, layout = onecolumn]{achemso}

\usepackage[version=3]{mhchem} 
\usepackage{graphics}
\usepackage{mathtools, amssymb, lipsum, nccmath}
\usepackage{float}

\usepackage{cuted}
\usepackage{booktabs}
\usepackage{{makecell}}
\usepackage{array}

\graphicspath{{figures/}}
\author{Ayush Jain}
\affiliation[GTMSE]
{School of Materials Science and Engineering, Georgia Institute of Technology, Atlanta, GA 30332, USA}
\alsoaffiliation[GTCoC]
{College of Computing, Georgia Institute of Technology, Atlanta, GA 30332, USA}
\author{Rishi Gurnani}
\affiliation[GTMSE]
{School of Materials Science and Engineering, Georgia Institute of Technology, Atlanta, GA 30332, USA}
\author{Arunkumar Rajan}
\affiliation[GTMSE]
{School of Materials Science and Engineering, Georgia Institute of Technology, Atlanta, GA 30332, USA}
\author{H.Jerry Qi}
\affiliation[GTME]
{School of Mechanical Engineering, Georgia Institute of Technology, Atlanta, GA 30332, USA}
\author{Rampi Ramprasad}
\affiliation[GTMSE]
{School of Materials Science and Engineering, Georgia Institute of Technology, Atlanta, GA 30332, USA}
\email{rampi.ramprasad@mse.gatech.edu}

\title[An \textsf{achemso} demo]
  {A Physics-Enforced Neural Network to Predict Polymer Melt Viscosity}

\abbreviations{IR,NMR,UV}
\keywords{American Chemical Society, \LaTeX}

\begin{document}

\begin{abstract}

Achieving superior polymeric components through additive manufacturing (AM) relies on precise control of rheology. One key rheological property particularly relevant to AM is melt viscosity ($\eta$). Melt viscosity is influenced by polymer chemistry, molecular weight ($M_w$), polydispersity, induced shear rate ($\dot\gamma$), and processing temperature ($T$). The relationship of $\eta$ with $M_w$, $\dot\gamma$, and $T$ may be captured by parameterized equations. Several physical experiments are required to fit the parameters, so predicting $\eta$ of a new polymer material in unexplored physical domains is a laborious process. Here, we develop a Physics-Enforced Neural Network (PENN) model that predicts the empirical parameters and encodes the parametrized equations to calculate $\eta$ as a function of polymer chemistry, $M_w$, polydispersity, $\dot\gamma$, and $T$. We benchmark our PENN against physics-unaware Artificial Neural Network (ANN) and Gaussian Process Regression (GPR) models.  Finally, we demonstrate that the PENN offers superior values of $\eta$ when extrapolating to unseen values of $M_w$, $\dot\gamma$, and $T$ for sparsely seen polymers.

\end{abstract}

\section{Introduction}

Additive Manufacturing (AM) enables the rapid creation of metal or polymer parts with previously-unimaginable features and topologies and is therefore poised to disrupt a variety of industries  \cite{AMML_prespectives, rheoinAM}. For polymers, achieving desired properties in the final component is determined by the appropriate choices of material chemistries with suitable rheological properties, as well as conditions adopted during the AM process such as temperature, extrusion rates, etc. At present, a limited palette of chemistries, properties, and conditions is utilized, generally guided by experience, intuition, and empiricism.

In this contribution, we adopt an informatics approach relevant to AM across the chemical and process condition space, to predict one critical rheological property of polymers, namely, the melt viscosity $\eta$. Informatics approaches have made major inroads in recent years within materials research \cite{chen2021polymer, rampiNRM, tran2024nrm}, leading to accelerated means for property predictions and providing guidance for the design of new materials \cite{PG, PGdielectric, retrosynth, polyg2g, liu2022prediction}. These methods start with available materials data on properties of interest. The materials are then represented numerically to capture and encode their essential features in a machine-readable format. The numerical representations, or fingerprints, are then mapped to available property data using machine learning (ML) algorithms, ultimately producing predictive models for the property considered \cite{PG, PGdielectric2, kuenneth2021copolymer, moriwaki2018mordred, polygnn, gurnani2024ai}. Within the AM space, similar methods have been used for final component property prediction \cite{AMmechprop}, process monitoring \cite{delli2018automated}, geometric configuration \cite{jin2020machine}, composition optimization \cite{jain2024machine}, and optimization of printing parameters (albeit mainly for powder-bed AM \cite{jin2020machine, AMML_prespectives}, but not as much for polymer melt extrusion AM \cite{rheoinAM}). Extrusion AM relies on the precise control of polymer melts, which currently requires data from extensive rheological experiments for each new chemistry. This is a bottleneck in the ink development process \cite{rheoinAM}. Therefore, predictive capabilities for rheological properties, such as $\eta$, are useful to reduce the number of physical experiments aimed at optimization and design. 

\begin{figure*}[!ht]
    \centering
    \includegraphics[scale=0.75, trim={0 0 0 0},clip]{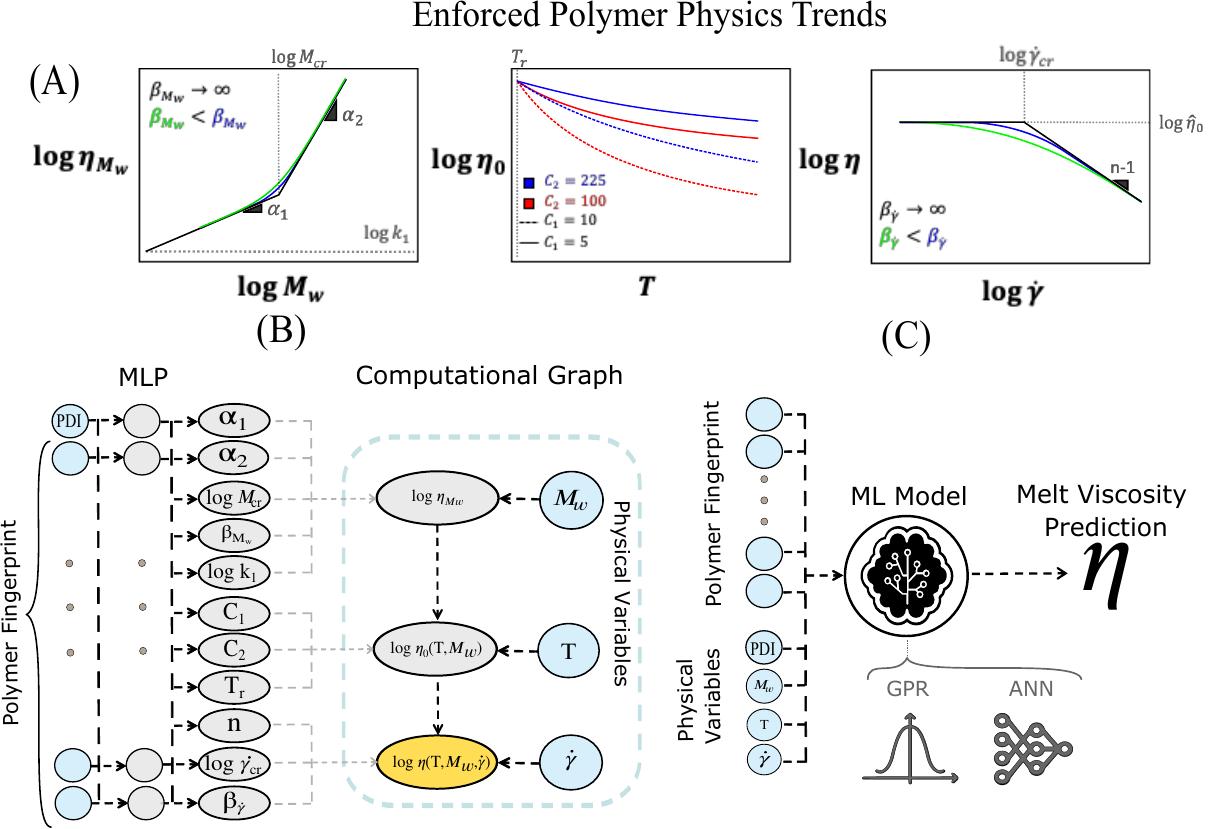}
    \caption{The melt viscosity ($\eta$) learning problem and machine-learning workflow. (A) Depictions of the functions used to describe the behavior of $\eta$ with respect to temperature ($T$), molecular weight ($M_w$), and shear rate ($\dot{\gamma}$). The functions are parametrized by empirical parameters with physical significance, elaborated in Table \ref{tbl:PENN_const} and in the Methods section. The $\eta$ dependence on $M_w$ is given by $\log {\eta}_{M_w}$ (Equation \ref{eq: Mw_smooth} in the Methods section). Empirical parameters define the slopes of the relationship at low $M_w$ ($\alpha_1$) and high $M_w$ ($\alpha_2$), the critical molecular weight ($M_{cr}$), the y-intercept of $\eta_{M_w}$ ($k_1$), and the rate of transition from low to high $M_w$ regions ($\beta_{M_w}$). The $\eta$ dependence on $T$ and $M_w$ is given by $\log \eta_0(T, M_w)$ (Equation \ref{eq: T} in the Methods section), and is parameterized by reference temperature ($T_r$) and empirical fitting parameters ($C_1$ and $C_2$). The effects of $C_1$ and $C_2$ are visualized by comparing the trends with different sampled values.   The $\eta$ dependence on $\dot\gamma$ is given by $\log \eta(T, M_w, \dot{\gamma})$ (Equation \ref{eq: shear_smooth} in the Methods section). The relevant parameters include shear thinning slope ($n$), the critical shear rate ($\dot\gamma_{cr}$), and the rate of transition from $\eta_0$ to shear thinning ($\beta_{\dot\gamma}$). (B) The Physics-Enforced Neural Network (PENN) architecture starts with an input containing the polymer fingerprint and the PDI. A Multi-Layer Perceptron (MLP) uses the concatenated input to predict the empirical parameters. Next, the computational graph uses the predicted empirical parameters to calculate $\eta$, via the encoded $\log {\eta}_{M_w}$, $\log \eta_0(T, M_w)$, and $\log \eta(T, M_w, \dot{\gamma})$ functions. The physical condition variables $\log M_w$, $\log \dot{\gamma}$ and $T$ are input to their respective functions. (C) Physics unaware Artificial Neural Network (ANN) and a Gaussian Process Regression (GPR) are baselines to compare with the PENN model. The input features to the ANN and GPR models are the concatenated polymer fingerprint, $T$, $M_w$, $\dot{\gamma}$, and PDI.}
    \label{fig:workflow}
\end{figure*}

Melt viscosity of polymers, beyond being a critical property, is attractive to model with ML because there is a reasonable amount of related literature data, although with limited chemical diversity compared to other polymer property datasets \cite{kuenneth2021copolymer, shukla2023polymer, tran2024nrm, gurnani2024ai, phan2024gas}. Additionally, there are known physical equations (albeit with empirical parameters) that describe the dependence of $\eta$ on its governing conditions: temperature (T), average molecular weight ($M_w$), and shear rate  ($\dot\gamma$) (Figure \ref{fig:workflow}A). For instance, it is known that $\eta$ increases with increasing $M_w$ (via piece-wise power law dependencies), decreases (non-linearly) with increasing $\dot\gamma$, and decreases (exponentially) with increasing T. Explicit functional forms and additional background on the behaviors are provided in the Methods section. Molecular weight distributions, quantified by the polydispersity index (PDI), are also known to affect melt viscosity \cite{pdi-effects, pdi-shear-effects, pdi-model}. With this situation in mind, previous works have also addressed the modeling of $\eta$ using ML \cite{polyviscML, silicatemeltsML, polysulfblendML, polyrheomodel}. While promising, the majority of these works have focused on specific scenario or are shown to predict unphysical results, making them difficult to apply.

In the present work, we create a physics-enforced neural network (PENN) framework that produces a predictive model of polymer melt viscosity which explicitly encodes the known physical equations while also learning the empirical parameters for new chemistries directly from available data. Physics-informed ML frameworks have shown great promise recently in many application spaces, including atomic modeling, chemistry-informed materials property prediction, and Physics Informed Neural Networks (PINNs) that solve partial differential equations \cite{atom_pinn_rampi,  digitalpinn, cheminfo,weightbiaspinn, PINN}. Our PENN for polymer melt viscosity prediction involves a Multi-Layer Perceptron (MLP) that takes as input the polymer chemistry (fingerprinted using our Polymer Genome approach \cite{PG}) along with the PDI of the sample, and predicts the empirical parameters as a latent vector (listed in Table \ref{tbl:PENN_const}), used to estimate $\eta$ as a function of $T$, $M_w$, and $\dot\gamma$. A computational graph then encodes the dependence of $M_w$, $\dot\gamma$, and $T$ on $\eta$ (see Figure \ref{fig:workflow}A) using the equations described in the Methods section. The entire framework (Figure \ref{fig:workflow}B) is trained on our dataset (elaborated in the Dataset section). The detailed architecture of this framework is described in the Methods section.

\begin{table}[!ht]
\small
  \begin{center}
  \begin{tabular}{p{15mm} p{100mm} p{20mm}}
    \hline
    Parameter & Physical Representation & Relevant Equation(s) \\
    \hline
    $C_1$  & Empirical Parameter for the $\eta$ - T relationship & 5 \\
    $C_2$  & Empirical Parameter for the $\eta$ - T relationship & 5\\
    $T_r$  & Reference temperature for the $\eta$ - T & 5\\
    $M_{cr}$  & Critical $M_w$, associated with the onset of polymer chain entanglement & 6,7,8 \\
    $\alpha_1$  & Slope of zero-shear viscosity ($\eta_0$) vs. $M_w$ when $M_w < M_{cr}$ (approximately 1) & 6,7,8 \\
    $\alpha_2$  & Slope of $\eta_0$ vs. $M_w$ when $M_w > M_{cr}$ (approximately 3.4) & 6,7,8\\
    $\beta_{M_w}$  & Measure of transition from $\alpha_1$ to $\alpha_2$ at $M_{cr}$ & 8\\
    $k_{1}$   & $\eta_0$ when $M = 0$ and $T= T_r$ & 6,7,8 \\
    $\dot{\gamma}_{cr}$  & Critical Shear Rate when $T= T_r$, associated with the onset of shear-thinning & 2,3,4 \\
    $n$ & Slope of shear thinning (typically 0.2-0.8 for polymer melts) & 2,3,4 \\
    $\beta_{\dot{\gamma}}$  & Measure of transition from zero-shear to shear-thinning & 4\\
    \hline
  \end{tabular} 
  \caption{Definitions of empirical parameters predicted by the Physics Enforced Neural Network (PENN) and their relevance to temperature ($T$), molecular weight ($M_w$), and shear rate ($\dot\gamma$) when calculating melt viscosity ($\eta$). The relevant equations in the Methods section are provided for each parameter.}
  \label{tbl:PENN_const}
  \end{center}
\end{table}

We find that this strategy is critical to obtain results that are physically meaningful in extrapolative regimes (e.g., ranges of T, $M_w$ and $\dot{\gamma}$ where there is no training data for chemistries similar to the queried new polymer). This ability is vital given our benchmarking dataset's sparsity, containing only 93 unique repeat units, although the total number of datapoints is 1903 (including $T$, $M_w$, $\dot\gamma$, and composition variations. As baselines to assess this PENN, we trained artificial neural network (ANN) and Gaussian process regression (GPR) models without any physics encoded. We find that the PENN model is more useful in obtaining credible extrapolative predictions. Our results indicate that informatics-based data-driven and physics-enforced (when possible) strategies can aid and accelerate extrusion AM innovations in sparse data situations.

\section{Results and Discussion}

\subsection{Dataset}
\begin{figure*}[ht!]
    \centering
    \includegraphics[scale=0.8]{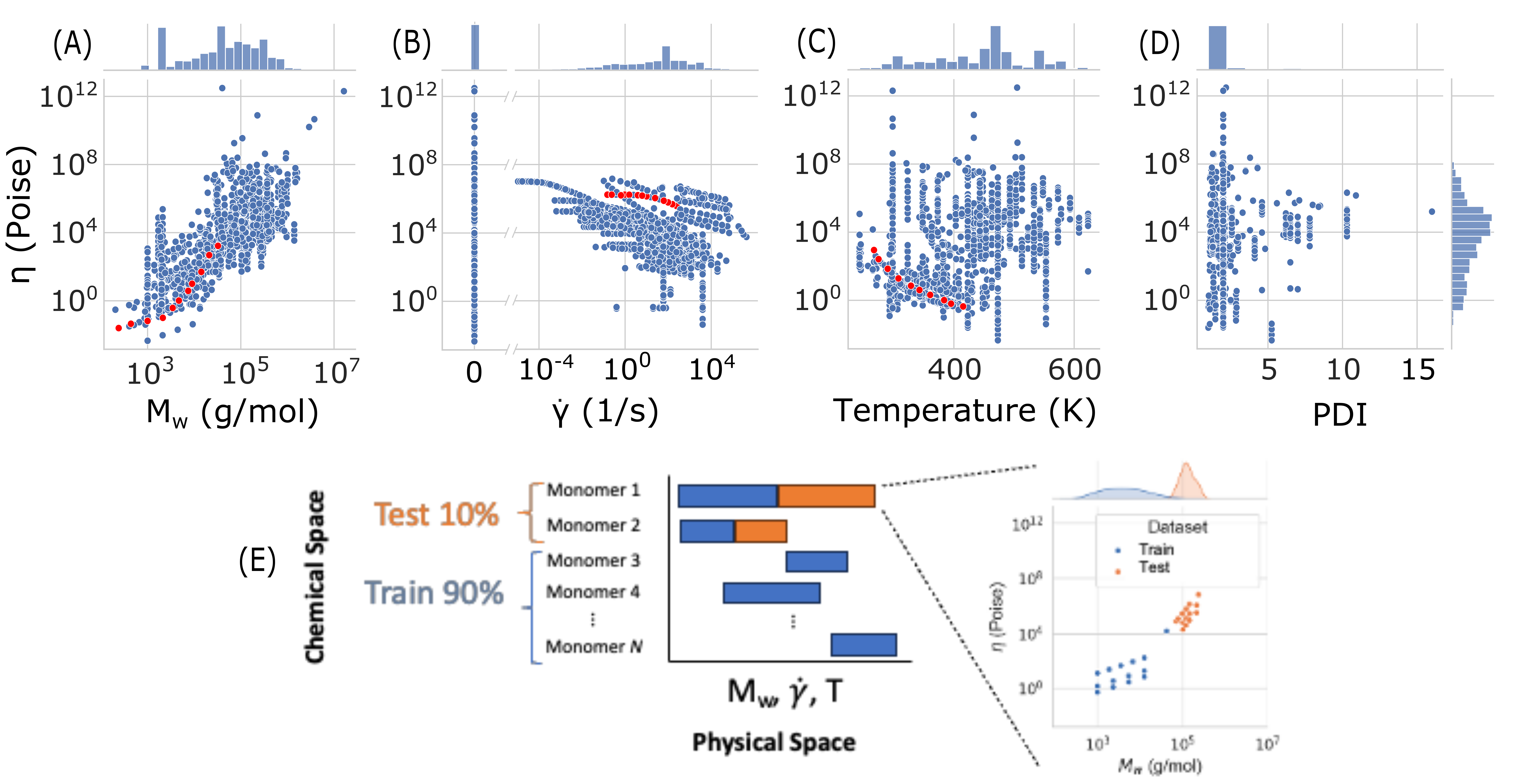}
    \caption{The joint distributions of A) molecular weight ($M_w$), B) shear rate ($\dot\gamma$), C) temperature ($T$), and D) polydispersity index (PDI) with respect to melt viscosity ($\eta$) are presented. The single distributions for the physical conditions are given on the top axes and the distribution of $\eta$ is given on the right-most axis. Each subplot contains all 1903 datapoints from the dataset. A, B, and C have highlighted samples in red that exemplify the dependencies depicted in Figure \ref{fig:workflow}A. E) Visual depiction of train-test splitting across chemical space and physical spaces for \textit{N} monomers in the dataset.}
    \label{fig:dataset}
\end{figure*}

Melt viscosity data was collected from the PolyInfo repository \cite{polyinfo} and from the literature cited by PolyInfo. Cited literature data was extracted from tables and figures with the help of the WebPlotDigitizer tool \cite{webplotdigitizer}. The final dataset shown in Figure \ref{fig:dataset} includes a total of 1903 datapoints composed of 1326 homopolymer datapoints, 446 co-polymer datapoints, and 113 miscible polymer blend datapoints. The dataset spans a total of 93 unique repeat units with variations in $M_w$, $\dot\gamma$, $T$, and PDI. For datapoints without a recorded PDI, we impute 2.06, the median PDI of the dataset. 

We found that $\eta$ at low $M_w$ were underrepresented when compared to $\eta$ measurements at high $M_w$. Using the zero-shear viscosity ($\eta_0$) relationship with $M_w$ (Figure 1A), we added 126 datapoints at low $M_w$ (included in the 1903 datapoints). This was achieved by identifying polymer chemistries with more than five $\eta_0$ datapoints at high $M_w$ and a recorded $M_{cr}$\cite{handbook-polymers-2nd}. Equation \ref{eq: Mw_piece} (Methods Section) was fit to each chemistry and extrapolated to estimate $\eta$ values at low $M_w$.

Because the viscosity values span several orders of magnitude (Figure \ref{fig:dataset}), we use the Order of Magnitude Error (OME) to assess ML model accuracy. OME is calculated by taking the Mean Absolute Error of the logarithmically scaled $\eta$ values. Models with lower OME exhibit more accurate predictions.

\subsection{Overall Assessment of Physical Intuition with Sparse Chemical Knowledge}

\begin{figure*}[!ht]
    \centering
    \includegraphics[width=\columnwidth]{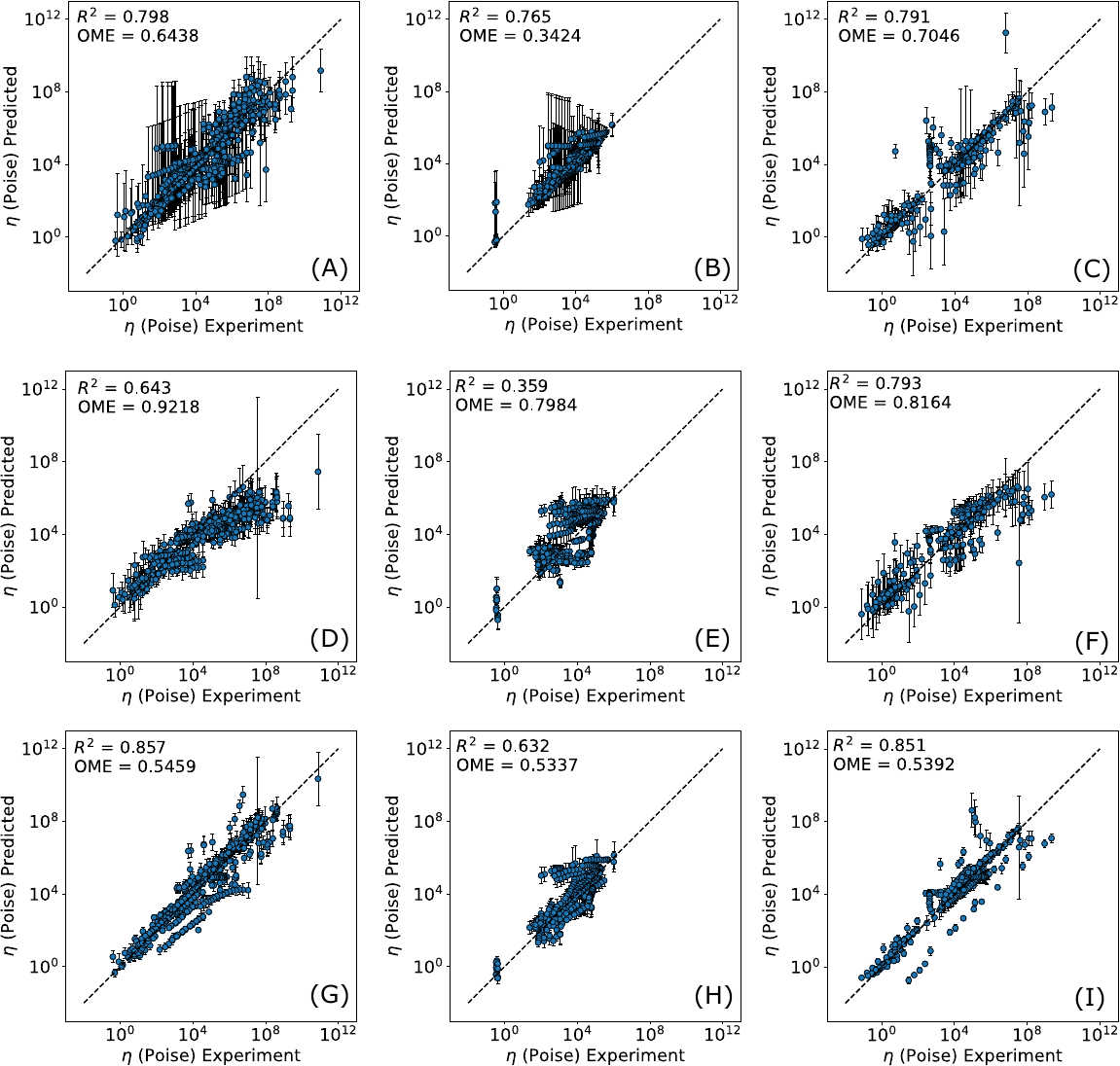}
    \caption{Parity plots are used to assess the models' overall predictive capabilities in new physical regimes based on the physical variable split for molecular weight ($M_w)$, shear rate ($\dot\gamma$), and temperature $T$. Results are compared between Gaussian Process Regression (GPR), Artificial Neural Network (ANN), and  Physics Enforced Neural Network (PENN) models. Each plot compares experimental values for melt viscosity ($\eta$) to the predicted $\eta$ across 3 unique test-train splits for each physical variable. The top row (A-C) contains GPR results for A) the $M_w$ split, B) the $\dot\gamma$ split C) the $T$ split. The middle row (D-F) contains ANN results for D) the $M_w$ split, E) the $\dot\gamma$ split F) the $T$ split. The bottom row (G-I) contains PENN results for D) the $M_w$ split, E) the $\dot\gamma$ split F) the $T$ split. The dotted black lines represent perfect predictions. The coefficient of determination ($R^2$) and Order of Magnitude Error (OME) are reported over these test sets.}
    \label{fig:parity}
\end{figure*}

An important future use case of our ML models is to estimate the melt viscosity in new physical regimes, given a small amount of knowledge of a given polymer and other chemistries. For example, given a few costly tests of a new polymer at a few molecular weights, one should be able to predict the viscosity at remaining molecular weights, and, likewise, across different shear rates and temperatures. Figure \ref{fig:dataset}E depicts how this ability was tested through a unique splitting of data into test/train sets across the chemical and physical regimes. First, the monomers were split into train (90\%) and test (10\%) sets. Within the test monomers, the median of the distributions of the test monomers with respect to a variable in the physical space was calculated. The median was used to split all datapoints containing that monomer: half for a final test split, and the other half for training. The upper or lower half going to testing was randomly chosen. This approach ensures that all the test data focuses on predicting in new physical regimes given a sparse amount of monomer data. This process was repeated three times for each of $M_w$, $\dot\gamma$, and $T$ to ensure that diverse tests were used for evaluation.


Figure \ref{fig:parity} shows the combined results of three trials for splits across all three physical variables. SI Section 1 shows parity plots that specify the results from each trial. The GPR, ANN, and PENN predictions have acceptable OMEs, indicating that all three can capture some chemical information and physical trends. The PENN results in a distinct decrease in OME (an average of 35.97\% improvement), and an increase in $R^2$ (up to 79\% for the $\dot\gamma$ split) from the ANN. The PENN also outperforms the GPR for the $M_w$ and $T$ splits, but the GPR is more accurate on the test set for $\dot\gamma$. In further analysis, we show how the physical viability of these predictions is scrutinized beyond the high-level trends of the parity plot.

\subsection{Distribution of Predicted Empirical Parameters}

\begin{figure*}[ht!]
    \centering
    \includegraphics[width=\columnwidth]{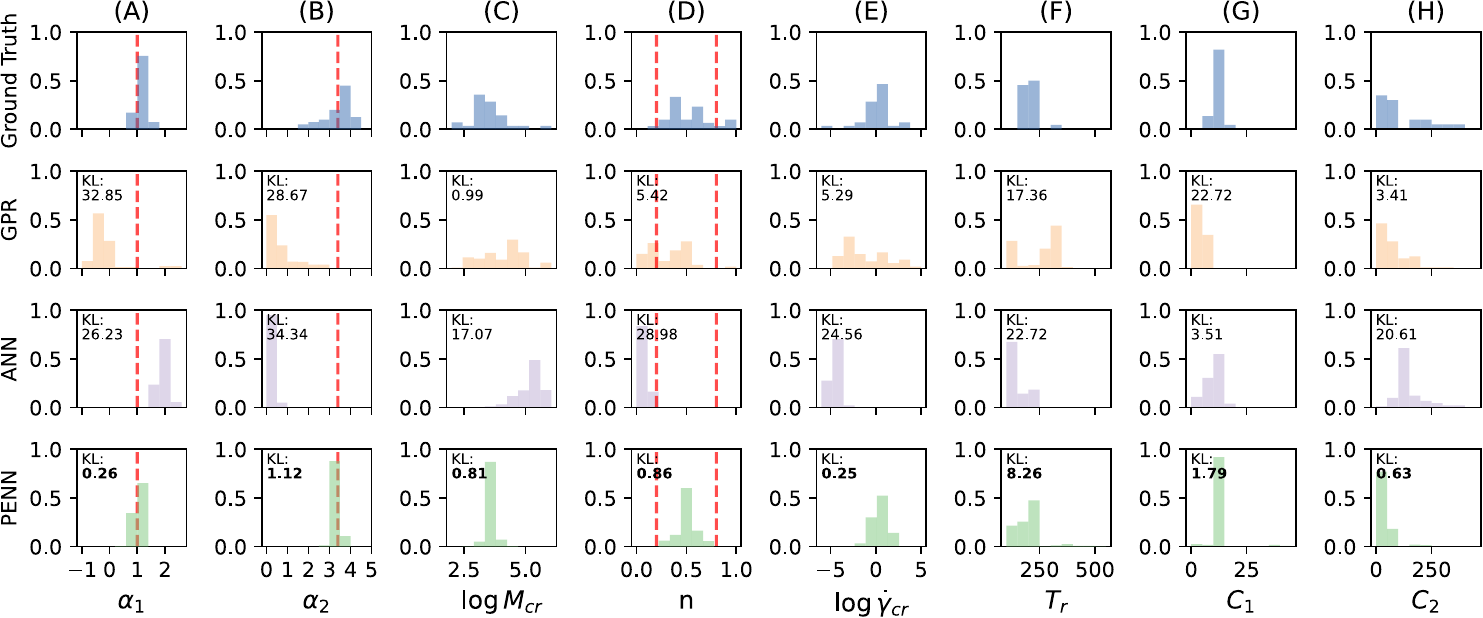}
    \caption{Normalized distributions of empirical parameter values found in the dataset (Ground Truth) are compared to parameter values predicted by Gaussian Process Regression (GPR), Artificial Neural Network (ANN) and Physics Enforced Neural Network (PENN) models. Each column compares a different parameter for the melt viscosity ($\eta$) relationship with molecular weight ($M_w$), shear rate ($\dot\gamma$), and temperature ($T$). The examined parameters include: A) $\alpha_1$, the slope of zero-shear viscosity ($\eta_0$) vs. $M_w$ correlation at low $M_w$ (accepted value of 1 depicted by the red dashed line) B) $\alpha_2$, the slope of $\eta_0$ vs. $M_w$ at high $M_w$ (accepted value of 3.4 depicted by the red dashed line), C) critical molecular weight ($M_{cr}$), D) $n$, the rate of shear thinning (accepted range of 0.2-0.8 depicted by the dashed red lines), E) critical shear rate ($\dot\gamma_{cr}$), F) reference temperature ($T_r$) of a polymer. G and H) show distributions for the $C_1$ and $C_2$ fitting parameters for the $\eta$-$T$ trend. The ground truth distributions represent 41 samples for $M_w$ parameters, 33 samples for $\dot\gamma$ parameters, and 22 samples for $T$ parameters. The Kullback–Leibler (KL) divergence of the model estimation distributions from the ground truth is given in the top left of each histogram. The lowest KL divergence among the three models is bolded for each parameter.}
    \label{fig:phys_const}
\end{figure*}

Despite the high overall performance of all three models, only the PENN model can produce physically credible predictions in regimes with restricted and sparse data. A comparison of the GPR, ANN, PENN models in estimating crucial empirical parameters (found in Table \ref{tbl:PENN_const}) from sparse data in the held-out set is detailed in Figure \ref{fig:phys_const}. 

To establish a benchmark for comparing the three models, we obtained ground truth values of the parameters from the dataset. We did this by identifying subsets of our dataset involving the same polymer with measured $\eta$ of several $T$, $M_w$, or $\dot\gamma$. If a subset contained at least five points, we fitted the corresponding equation (Equation \ref{eq: T}, \ref{eq: Mw_piece}, or \ref{eq: mod_cross_yas}) to obtain empirical parameters. The distributions of these ground truth parameter values are shown in the first row of Figure \ref{fig:phys_const}. There are a limited number of ground truth values because a small number of datapoints satisfy the above conditions. Nevertheless, this small sample set allowed us to make a few inferences about expected viscosity trends. The ground truth values of $\alpha_1$ and $\alpha_2$ are close to the theoretical values of 1 and 3.4, respectively\cite{Mw-background} (background provided in Methods). $\alpha_2$ values were occasionally less than the expected 3.4, possibly due to outliers or errors in fitting a small number of datapoints. The fitted $\log M_{cr}$ values fell within a range of $10^{2.5} - 10^{5}$ g/mol. For shear parameters, the majority of samples are found to have $n$ in a range of $0.2-0.8$, which is typical for polymer melts \cite{PolyReochap5}.  The obtained $\dot{\gamma}_{cr}$ values were found in the range of $10^{-3}-10^4$ 1/s. The fitted $T_r$ values are mostly in a range of $T_r < 250 K$. This is low when compared to $T_g$ values found in thermal property datasets \cite{kuenneth2021copolymer}. In our dataset, the datapoints that could be fitted to the $\eta$-$T$ relationship were observed at $T < 475K$, so low $T_r$ values could be overrepresented in the ground truth. The $C_1$ parameter average was 11.8 and the $C_2$ parameter average was 159.42 K. This analysis of the ground truth data suggests the desired parameter values our models should predict.

We used two different methods to obtain parameter estimations from the models: one method is unique to the PENN model, and another approach for the purely data-driven ANN and GPR. The PENN model automatically predicts each of the empirical parameters (see Figure \ref{fig:workflow}B), which are used in the computational graph to predict $\eta$. The ANN and GPR do not directly predict the parameters, so we used a fixed extrapolation procedure. The procedure involved selecting an unseen data point and varying a physical variable (one of $M_w$, $\dot\gamma$, and $T$) within a predetermined range while holding the other two constant. The ranges for each variable encompass similar orders of magnitude as those present in the training dataset (Figure \ref{fig:dataset}). For $M_w$ extrapolation, a range of $10^2 - 10^7$ g/mol was used to encompass low and high $M_w$. For shear rate extrapolation, a range of $10^{-5} - 10^6$ 1/s was used to model behaviors in zero-shear and shear-thinning regimes. For temperature extrapolation, ranges of $\pm20$ K from the original data point's temperature were used to stay within the boundary constraints of Equation \ref{eq: T}. Using this procedure, sets of predictions were made on every unseen datapoint and fit using Equations \ref{eq: T}, \ref{eq: Mw_piece}, or \ref{eq: mod_cross_yas}, yielding estimated values of the empirical parameters. 


In Figure \ref{fig:phys_const}, we show the feasibility of the models' empirical parameter predictions evaluated against the ground truth values and accepted values (elaborated in the Methods section). For parameters where a theoretical value is well-defined, the Root Mean Square Error (RMSE) of the predictions' deviation from this value is calculated. The parameter prediction distribution is also compared to the ground truth distribution through a discrete Kullback–Leibler (KL) divergence,
\[
KL(P \parallel Q) = \sum_{i} P(i) \log \left(\frac{P(i)}{Q(i)}\right).
\]
Intuitively, the KL divergence is a measure of how one probability distribution $P$ deviates from a reference distribution $Q$ over a set of intervals $i$. A lower divergence indicates that the predicted parameter distribution is closer to the ground truth. The KL divergence was calculated by finding the entropy between the discretized probability distributions of the ground truth and the ML prediction. 


From Figure \ref{fig:phys_const}, it can be seen that GPR struggles to predict expected parameter values. The GPR predictions for $\alpha_1$ deviate from $1$ by an RMSE of 1.26. For some polymers, GPR predicts $\alpha_1 \leq 0$. The GPR predictions for $\alpha_2$ deviate from 3.4 by an RMSE of 2.87, and are significantly lower than the ground truth values in the dataset. Most predicted values for $\log M_{cr}$ are within the same range as the ground truth, but the proper low and high entanglement behavior is not captured which decreases the credibility of these fittings. For the shear thinning parameter $n$, some values fall within the expected range of $0.2-0.8$ \cite{PolyReochap5} for polymer melts, but others are closer to 0, indicating that the expected shear thinning behavior is not always predicted. The predicted $\dot\gamma_{cr}$ distribution is lower than the ground truth, indicating that the GPR forecasts the onset of shear-thinning at a significantly lower $\dot\gamma$ than observed (if shear thinning is predicted at all). On temperature dependence, some $T_r$ values are predicted higher than what is seen in the dataset.

The ANN's failure to capture correct physical trends is also evident in the distributions of its fitted parameters. The RMSEs for the ANN's estimated $\alpha_1$ and $\alpha_2$ values are 1.31 and 2.79, respectively. ANN overestimates $\alpha_1$ and underestimates $\alpha_2$ and therefore does not capture the effects of high $M_w$ chain entanglement. The ANN predictions estimate a low $n$ for a subset of polymers, which goes against the definition of shear thinning. The predicted $\dot\gamma_{cr}$ values are lower than the ground truth distribution, indicating that the ANN struggles to capture the shear-thinning transition region from the dataset. The ANN predictions for the $T$ trend are closest to the ground truth in comparison to its trends of the other variables, because $T$ is a smoother, exponential function (Figure \ref{fig:workflow}A), enabling an easier average fitting.

The PENN outperforms the ANN in estimating feasible empirical parameters as depicted by lower KL Divergence values in the last row of Figure \ref{fig:phys_const}. The RMSEs of the predicted $\alpha_1$ and $\alpha_2$ values are 0.05 and 0.17, which are substantially smaller than that of the ANN. Moreover, all the predicted values of $\log M_{cr}$ are within the ground truth range of $2.5 - 5$. The PENN model can also learn the correct shear thinning phenomenon by predicting $n$ values between $0.2-0.8$\cite{PolyReochap5} and a $\dot\gamma_{cr}$ distribution that mirrors the dataset. The PENN's predicted range of $T_r$ is closest to the ground truth. For the $C_1$ parameter, the PENN predicted distribution is closest to the proposed value of $C_1=7.60$ (detailed in the Methods section), also having the lowest divergence from the ground truth. For $C_2$ predictions, although the KL Divergence of the PENN is lower than the ANN, the PENN is confined to much lower values of $C_2$, and has an average much lower than some experimentally derived values, such as $C_2$ = 227.3 K \cite{WLF2}. 

Overall, the average KL divergence across all parameter distributions for the GPR, ANN, and PENN are 14.59, 22.24, and 1.74, respectively. The overall distributions of empirical parameters points to the PENN having greater capabilities for producing physically correct results, than a purely data-driven model.

\subsection{Performance in Extrapolative Regimes}

\begin{figure*}[ht!]
    \centering
    \includegraphics[scale=0.8]{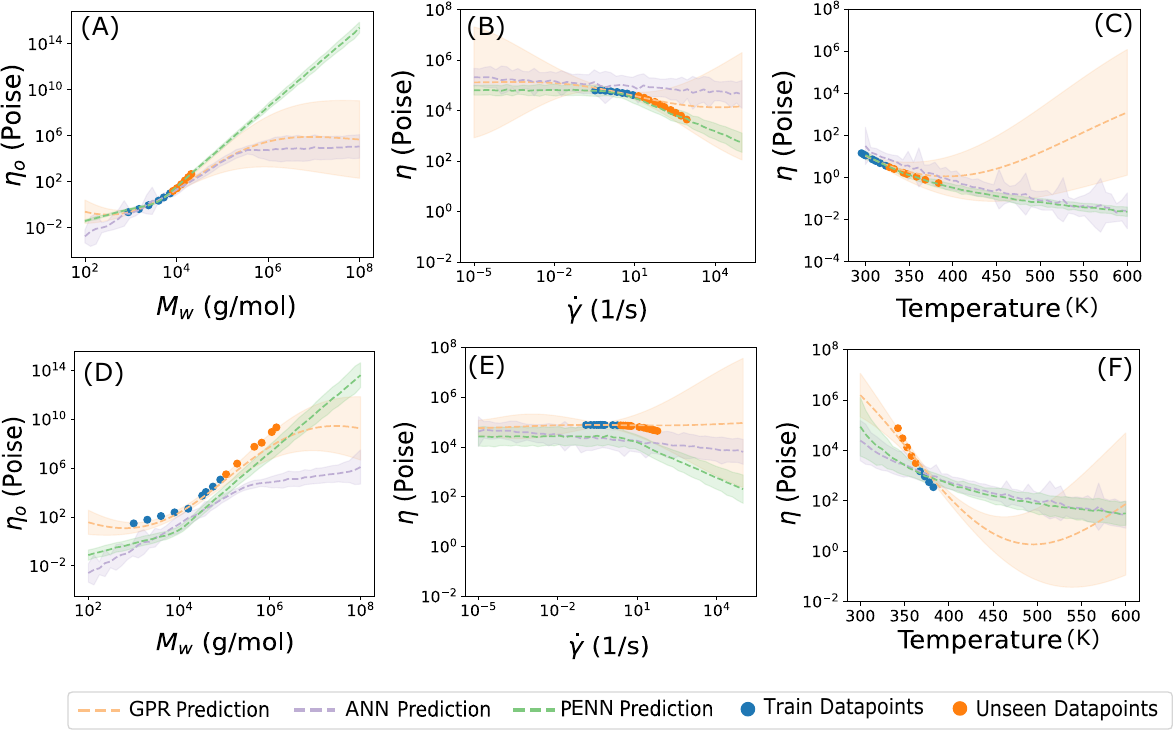}
    \caption{
    Examples of accurate (A-C) and inaccurate (D-F) melt viscosity ($\eta$) and zero-shear melt viscosity ($\eta_0$) predictions over wide ranges of molecular weight ($M_w$), shear rate ($\dot\gamma$), temperature ($T$) by the Physics Enforced Neural Network (PENN) models. The extrapolated predictions are compared to those by Gaussian Process Regression (GPR) and Artificial Neural Network (ANN) models given the same training information. A) is a good $\eta_0$-$M_w$ extrapolation for [*]CCCCCCCCCCOC(=O)CCCCC(=O)O[*] at $T=382.15$ K. B) is a good $\eta$-$\dot\gamma$ extrapolation for a copolymer of [*]CC([*])CC(C)C and [*]CC([*])CCCCCCCC (0.968:0.032) ($M_w = 290000$ g/mol, PDI = 7.8) at $T = 543.15$ K. C) is a good $\eta$-$T$ extrapolation for [*]CCOCCOCCOC(=O)CCCCCCCCC(=O)O[*] ($M_w$ = 2000 g/mol, $\dot{\gamma}$ = 60 1/s). D) is an unsuccessful $\eta_0$-$M_w$ extrapolation for [*]C=CCC[*] at $T = 490.15$ K, with possible mispredictions of $M_{cr}$ and $k_1$. E) is an unsuccessful $\eta$-$\dot\gamma$ extrapolation for a copolymer of [*]C[*] and [*]CC([*])OC(C) (0.72:0.28) ($M_w = 60000$ g/mol), with possible misprediction of $\hat \dot\gamma_{cr}$ and $\eta_0$. F) is an unsuccessful $\eta$-$T$ extrapolation for [*]CC(O)COc1ccc(C(C)(C)c2ccc(O[*])cc2)cc1 ($M_w$ = 1696 g/mol, $\dot{\gamma}$ = 0.0 1/s) with a possible misprediction of $T_r$.}
    \label{fig:extraps}
\end{figure*}

In Table \ref{tbl:success_predict}, we summarized the performance of predicted $\eta$ profiles over wide ranges of $M_w$ (256 extrapolations), $\dot\gamma$ (71 extrapolations), and $T$ (127 extrapolations) for all three models considered. We define a successful extrapolation as a model being able to predict the correct trends while maintaining accuracy over the train and test points. Overall, the PENN successfully predicts 80.4\% of $M_w$ extrapolations, 49.2\% of $\dot\gamma$ extrapolations, and 54.1\% of $T$ extrapolations. The ANN rarely achieves correct physical trends for $M_w$ or $\dot\gamma$ extrapolations in the span of the dataset and only predicts successful profiles for 17.2\% of $T$ extrapolations. The GPR model also exhibits a low performance in extrapolation. There are several instances (given in the brackets in Table \ref{tbl:success_predict}) where the ANN and GPR successfully fit the data points but fail to extrapolate correctly beyond the dataset. This underscores the need for information beyond experimental data to enable extrapolation to new physical regimes.

\begin{table}[!ht]
\small
  \begin{center}
  \begin{tabular}{p{10mm} | p{10mm}p{10mm}p{10mm}}
    \hline
    Model & $M_w$ & $\dot\gamma$ & $T$ \\
    \hline
    PENN & \textbf{80.4\%} & \textbf{49.2\%} &\textbf{54.1\%} \\
    ANN & 4.30\% [64.4\%] & 4.22\% [19.7\%] & 17.2\% [19.5\%] \\
    GPR & 0.0\% [68.0\%] & 7.04\% [40.8\%] & 7.03\% [28.1\%] \\
    \hline
  \end{tabular} 
  \caption{Extrapolative predictive performance of the PENN, ANN, and GPR models along the unseen molecular weight ($M_w$), shear rate ($\dot\gamma$), and temperature $T$ regimes. The values within brackets for the ANN and GPR show the percentages of extrapolations where the unseen data was predicted correctly, but the extrapolated trend beyond available data regimes was physically incorrect.}
  \label{tbl:success_predict}
  \end{center}
\end{table}


Figure \ref{fig:extraps} shows a few examples of the extrapolation results summarized in Table \ref{tbl:success_predict}. A much larger set of examples of both successful and unsuccessful extrapolations by the PENN compared to the GPR and the ANN are provided in SI Section 2. Figure \ref{fig:extraps}A-C shows examples of the PENN correctly extrapolating $\eta$ given a small amount of information about a monomer in another part of the physical regime in unseen regimes. The ANN and GPR models are uncertain in these unseen regimes, resulting in large confidence intervals. In Figure \ref{fig:extraps}A, the PENN model accurately predicts the region near $M_{cr}$ where the $\eta$-$M_w$ relationship transitions from unentangled to entangled, and can therefore accurately predict $\eta$ values at high $M_w$, despite not having seen any data in this region. The errors for the ANN and GPR in Figure \ref{fig:extraps}A are low, within approximately an order of magnitude of error. However, the ANN predictions have a near-constant slope around $M_{cr}$ (implying $\alpha_1 \approx \alpha_2$) and are inconsistent with the effects of polymer chain entanglements at high $M_w$. The GPR model also fails to predict a higher $\alpha_2$ slope. In Figure \ref{fig:extraps}B, only the PENN model predicts a zero-shear and shear-thinning region when predicting the $\eta$-$\dot\gamma$ relationship of the given copolymer. The GPR model fits the training points but mispredicted shear-thinning at high shear rates. The ANN model predicts a decreasing relationship consistent with shear-thinning but doesn't predict the zero-shear region.  This could be an example of spectral bias within neural networks \cite{rahaman2019spectral}, where the general decreasing trend of $\eta$-$\dot\gamma$ is "low frequency" and is captured by the ANN. In contrast, the transition regions are of a "higher frequency" and are not captured by the ANN. In Figure \ref{fig:extraps}C, the PENN model predicts the correct $\eta$-$T$ relationship. The ANN model also predicts an exponential relationship but with a higher inaccuracy. The GPR model fits both the training and unseen datapoints, but predicts an unphysical trend beyond this. Overall, the PENN model makes predictions that follow the expected behaviors (Figure \ref{fig:workflow}A)  of polymer melts.

Correctly extrapolated samples by the PENN model, such as the ones in Figure \ref{fig:extraps}A-C make up 67.5\% of the extrapolated test cases, which is a significant improvement relative to both the ANN and GPR. The PENN model also has room for improvement, especially when applied to datasets with low chemical diversity. Overfitting to a small set of chemistries in training can lead to the inaccurate prediction of parameters when making predictions for unseen chemistries. This behavior is demonstrated in Figure \ref{fig:extraps}D-F, where the PENN predicts a plausible rheological trend but incorrect values for unseen polymers. However, the PENN model introduces a layer of interpretability unavailable to physics-unaware models. Based on the predictions we can reasonably infer which parameters were over- or under-estimated. In Figure \ref{fig:extraps}D, the PENN model predicts near-correct $\alpha_1$ and $\alpha_2$ slopes, but the predicted $M_{cr}$ and $k_1$ values are underestimated. Figure \ref{fig:extraps}E depicts how an underestimated $\eta_0$ (caused by inaccuracies in predicted $M_{cr}$, $\alpha_1$, $\alpha_2$ and/or $k_1$) can cause inaccurate $\eta$ predictions for all other $\dot \gamma$ values. We also see this phenomenon in Figure \ref{fig:extraps}F, where $T_r$ is likely underestimated. The propagating error causes the PENN model to predict an inaccurate trend across the entire spectrum of $T$. Despite these errors, the pinpointing of the PENN's weak spots can be used to add targeted training data to improve the model. This level of interpretation is unique to the PENN and cannot be done for the GPR and ANN.

These examples of extrapolations provide insights into the applicability of PENN versus pure data-driven methods when using datasets that contain limited chemistries. The equations used in the PENN are based on assumptions and generalizations, and may not account for all physical nuances. These must be considered when applying PENNs to future material design and process optimization problems.

\section{Conclusion}

In this study, we introduce a Physics Enforced Neural Network (PENN), a strategy that combines data-driven techniques with established empirical equations, to predict the melt viscosity of polymer melts with better physics-guided generalization and extraplation. The PENN makes predictions across many chemical compositions and relevant physical parameters, including molecular weight, shear rate, temperature, and polydispersity index. We compared our PENN approach against the purely data-driven, physics-unaware, Artificial Neural Network and Gaussian Process Regression. In extrapolative regimes, our PENN model outperforms the physics-unaware counterparts and offers an elevated level of interpretability and generalizability. To enhance generalizability across chemistries, future work could increase the chemical space in the dataset through new experiments, molecular dynamics simulations, and/or more aggressive data acquisition from literature.

This work has profound implications for additive manufacturing (AM) and materials informatics. The PENN model's capability to guide the rheological control of diverse polymer resins accelerates the development of new printing materials, thereby expanding AM's utility. Our methodology offers a blueprint for modeling other properties governed by empirical equations. The initial success of the PENN architecture for melt viscosity is a powerful demonstration of how data-driven insights combined with established knowledge can propel us into a new era of rapid advancements in materials science and engineering.

\section{Methods}

\subsection{Fingerprinting and Feature Engineering}
The chemical attributes of a polymer are represented by a unique fingerprinting scheme. The fingerprints (FPs) contain features derived from atomic-level, block-level, chain-level, and morphological descriptors of a polymer as described at length earlier\cite{PG}. The dataset contains homo- and co-polymers, and miscible polymer blends. Co-polymers and blends contain multiple repeating units, each with a separate FP. For co-polymers, the FP of each unit was aggregated to a single copolymer FP using a weighted average (with weight equal to composition percentage) \cite{kuenneth2021copolymer}. Similar to previous work\cite{kuenneth2021copolymer}, all co-polymers were treated as random. For miscible polymer blends, the FP of each unit was aggregated to a single FP using a weighted harmonic average (with the weight equal to composition percentage) \cite{shukla2023polymer}. For blends containing units with different $M_w$ and/or PDI, the weighted average over each unit was used. 

\subsection{Enforced Polymer Physics Trends}

In this section, we detail the physics-based correlations included within the Physics Enforced Neural Network (PENN). 
We enforce dependencies of $\eta$ on temperature ($T$), molecular weight ($M_w$), and shear rate ($\dot\gamma$) through $\eta(M_w, T, \dot\gamma)$, which we derive below.  

\subsubsection{Preamble: Smoothing of Piecewise Functions}

When going from one function $g(a,b)$ in a low regime ($a<b$) to another function $h(a,b)$ in a high regime ($a>b$), we can use the smoothened Heaviside step function,

\begin{equation}
    \begin{array}{c}
        H_{\beta} = \frac{1}{1 + \exp(- \beta x)},
    \end{array}
    \label{eq: heaviside_smooth}
\end{equation}
where $\beta$ is a tunable rate of transition.

A function $f(a,b)$ that transitions from $g(a,b)$ to $h(a,b)$ is given by
\[
f(a,b) = g(a,b) \times H_{\beta}(b-a) + h(a,b) \times H_{\beta}(a-b)
\]

\subsubsection{$\eta$ dependence on $\dot\gamma$, $T$, and $M_w$}

The $\eta$ dependence on $\dot\gamma$ follows the physics of shear-thinning fluids \cite{shear-effects, polymermeltinfo, polyphysicstxtbook}. In these fluids, at low $\dot\gamma$, there is not enough force between chains to break entanglements and cause movement, so $\eta$ remains constant at $\eta_0$. At a critical shear rate, $\dot{\gamma}_{cr}$, the shear force is high enough to cause chain alignment, making chain diffusion easier. Beyond $\dot{\gamma}_{cr}$, $\eta$ decreases according to a shear-thinning linear power law. \cite{shear-effects}. This trend can be represented by a function (Equation \ref{eq: mod_cross_yas}) across both the zero-shear and shear thinning regimes \cite{Shaw-zero-shear, crossVcar, shear-effects, high-shear},

\begin{equation} 
\centerline{$\begin{aligned}
& \eta(M_w, T, \dot\gamma) = \frac{\eta_0(M_w,T)}{(1+\frac{\dot\gamma}{\dot\gamma_{cr}})^{1-n}} \\
& \log \eta = \log \eta_0(M_w,T) + (n-1) \log(1+\frac{\dot\gamma}{\dot\gamma_{cr}})
\end{aligned}$}
\label{eq: mod_cross_yas}
\end{equation}
where the parameter $n$ describes the sensitivity to shearing \cite{mastercurves}. For shear-thinning fluids, $n<1$. For most polymer melts, $n$ is empirically known to be in the range of $0.2-0.8$.\cite{PolyReochap5} 


Equation \ref{eq: mod_cross_yas} is unfavorable to use directly because $\dot\gamma$ spans several orders of magnitude, so $\log \dot\gamma$ must be used as an input. Equation \ref{eq: mod_cross_yas} cannot be adapted to use $\log \dot\gamma$ as an input (due to the +1 in the denominator), so we depict the relationship across the low $\dot{\gamma}$ and high $\dot{\gamma}$ regimes as a piecewise function on the log-scale,


\begin{equation}
\log \eta =
\begin{cases}
  \log \eta_0 & \text{if } \dot\gamma << \dot\gamma_{cr} \\
  \log \eta_0 + (n-1)\log(\frac{\dot{\gamma}}{\dot{\gamma}_{cr}}) & \text{if } \dot\gamma >> \dot\gamma_{cr} \\
\end{cases} 
\label{eq: shear_piece}
\end{equation}

We smooth Equation \ref{eq: shear_piece} with $H_{\beta_{\dot\gamma}}$ to get  $\log \eta(M_w, T, \dot\gamma)$ (Equation \ref{eq: shear_smooth}), 

\begin{align}
  \begin{aligned}
  & \medmath{\log {\eta}(M_w, T, \dot\gamma)
   = \log{\eta_0(M_w,T)} \times H_{\beta_{\dot\gamma}}(\log \dot{\gamma}_{cr} - \log \dot{\gamma})}
  \\[1ex]
    & \!\medmath{+ (\log \eta_0(M_w,T) + (n-1)\log(\frac{\dot{\gamma}}{\dot{\gamma}_{cr}})) \times H_{\beta_{\dot\gamma}}(\log \dot{\gamma} - \log \dot{\gamma}_{cr})},
  \end{aligned}
  \label{eq: shear_smooth}
\end{align}
where $\beta_{\dot\gamma}$ is a parameter that dictates the rate of shift from zero-shear to shear-thinning. For our implementation, we found that optimization over the $\dot\gamma$ domain was optimal when $\beta_{\dot\gamma} = 30$.

$\log \eta_0(M_w,T)$ is defined by the $T$ dependence. As temperature increases, so does the rate of molecular self-diffusion, resulting in lower $\eta$ seen in fluidic polymer melts \cite{polyphysicstxtbook}. The William-Landel-Ferry (WLF) equation\cite{WLF, WLF2} describes the exponential decrease in $\eta$ as the temperature increases. Therefore, we can encode temperature dependence as

\begin{equation} 
\begin{split}
\eta_0 &= \eta_{M_w} \times 10^{\frac{-C_1(T-T_r)}{C_2 + (T-T_r)}} , \forall T \geq T_r \\
\log \eta_0(M_w, T) &= \log \eta_{M_w} \times {\frac{-C_1(T-T_r)}{C_2 + (T-T_r)}}  , \forall T \geq T_r
\end{split}
\label{eq: T}
\end{equation}
where $T_r$ is a reference temperature and $C_1$ and $C_2$ are material-dependent empirical parameters. The values for these are dependent on polymer chemistry. $C_1$ = 7.60 and $C_2$ = 227.3 K are examples of values that have been proposed \cite{WLF2} from observations of experiments on a small subset of polymers. The reference temperature $T_r$ is within a few degrees of the glass transition temperature $T_g$. It has been proposed that the WLF relationship holds within the range of $T_g$ to $T_g + 200$K \cite{WLF2}. 

$\eta_{M_w}$ is defined by the $M_w$ dependence. Longer and heavier polymer chains experience increased entanglements, which hinder chain reptation in the polymer melt at low shear. \cite{Mw-background, polyphysicstxtbook}  Equation \ref{eq: Mw_piece} is a piece-wise power law that describes this phenomenon. 

\begin{equation}
\eta_{M_w} =
\begin{cases}
  k_1 M_w^{\alpha_1} & \text{if } M_w < M_{cr} \\
  k_2 M_w^{\alpha_2} & \text{if } M_w \geq M_{cr} \\
\end{cases}
\label{eq: Mw_piece}
\end{equation}
where 
\[
k_2 = k_1 M_{cr}^{\alpha_1 - \alpha_2}. 
\]

$M_{cr}$ is the critical molecular weight, above which entanglement density is high enough to increase the impact of $M_w$ on $\eta_0$. The two power laws intersect at $M_w=M_{cr}$ \cite{Mw-background}. $M_{cr}$ is found to be approximately 2-4 times the molecular weight at which chain entanglement starts, but the exact value is polymer dependent.\cite{polyphysicstxtbook} $\alpha_1$ is the slope of the $\log \eta_0$-$\log M_w$ curve if $M_w < M_{cr}$ and $\alpha_2$ is the slope if $M_w \geq M_{cr}$. Typically, $\alpha_1$ is theoretically and empirically determined to be about 1, while $\alpha_2$ is found to be about 3.4 \cite{Mw-background, polyphysicstxtbook}, but the exact value is dependent on the polymer. $k_1$ and $k_2$ are the y-intercepts of each power law and are polymer-dependent.

$M_w$ and $\eta_0$ span several orders of magnitude, so we use Equation \ref{eq: Mw_piece} in the log-scale to get Equation \ref{eq: Mw_log_piece}, 

\begin{equation}
\log \eta_{M_w} =
\begin{cases}
  \log k_1 + {\alpha_1} \log M_w & \text{if } M_w < M_{cr} \\
    \log k_1 + {(\alpha_1 - \alpha_2)} \log M_{cr} & \text{if } M_w \geq M_{cr} \\
  \quad + \alpha_2 \log M_w
\end{cases}
\label{eq: Mw_log_piece}
\end{equation}

Smoothing Equation \ref{eq: Mw_log_piece} with $H_{\beta_{M_w}}$ gives Equation \ref{eq: Mw_smooth}, \\

\begin{align}
  \begin{aligned}
  & \medmath{\log \eta_{M_w}}
  \\[1ex]
   & \!\medmath{= [\log k_1 + {\alpha_1} \log M_w] * H_{\beta_{M_w}}(\log M_{cr} - \log M_w) }
  \\[1ex]
    & \!\medmath{+ [\log k_1 + {(\alpha_1 - \alpha_2)} \log M_{cr}] * H_{\beta_{M_w}}(\log M_{w} - \log M_{cr})}, \\
  \end{aligned}
  \label{eq: Mw_smooth}
\end{align}
where $H_{\beta_{M_w}}$ is the smoothened Heaviside step function using $\beta_{M_w}$, a parameter which dictates the rate of shift from $\alpha_1$ to $\alpha_2$. 

Therefore, Equations \ref{eq: shear_smooth}, \ref{eq: T}, and \ref{eq: Mw_smooth} determine the $\log \eta(M_w, T, \dot\gamma)$. The predicted parameters $n$, $\dot\gamma_{cr}$, $\beta_{\dot\gamma}$ determine the $\dot\gamma$ dependence in $\log \eta(M_w, T, \dot\gamma)$, which is also a function of $\eta_0(M_w, T)$. The predicted parameters $C_1$, $C_2$, and $T_r$ determine the $T$ dependence in $\eta_0(M_w, T)$, which is also a function of $\eta_{M_w}$. The predicted parameters $\alpha_1$, $\alpha_2$, $M_{cr}$, $\beta_{M_w}$, and $k_1$ determine the $M_w$ dependence in $\eta_{M_w}$. The parameter outputs of the MLP have physically appropriate bounding ranges (elaborated in SI Section 3).

\subsubsection{$\eta$ dependence on $PDI$}

The dispersity of molecular weights in a polymer melt affects the bulk motion of polymer chains \cite{polyphysicstxtbook}. For example, a short and long chain may diffuse differently compared to two medium-sized chains. Therefore, using just the $M_w$ without any knowledge of dispersity can mislead the ML model. We account for dispersity by using the polydispersity index (PDI),
\[
PDI= \frac{M_w}{M_n},
\]

where $M_n$ is the number average molecular weight. Empirical models for this relationship \cite{pdi-effects, pdi-model} may require detailed information on the specific shape of the molecular weight distribution of a polymer melt. Not all of our data points contain proper information on PDI (as discussed in the Results section), so we do not directly encode $\eta$-PDI trends within the computational graph. Instead, the PDI could affect the transitions in the critical regimes of the $\eta_0$-$M_w$ relationship and the $\eta$-$\dot\gamma$ relationship (when $\dot{\gamma} = \dot{\gamma}_{cr}$ or $M_w = M_{cr}$) \cite{pdi-shear-effects, pdi-model, pdi-effects}. We incorporate this effect through the parameters $\beta_{M_w}$ and $\beta_{\dot\gamma}$ (described in Table \ref{tbl:PENN_const}). A higher value of $\beta_{M_w}$ or $\beta_{\dot\gamma}$ creates a quicker transition within their respective critical regimes.


\subsection{PENN Training}



This entire PENN architecture is trained, in part, to minimize the error of viscosity predictions. The sum of these errors across all $n$ training points is called the viscosity loss $\mathcal{L}_{\eta}$, defined in Equation \ref{eq: visc_loss}. Each data point is denoted by its index $i$.

\begin{equation}
    \mathcal{L}_{\eta} = \frac{1}{n} \sum_{i=1}^n (\hat{\eta}_i - \eta_i)^2
\label{eq: visc_loss}
\end{equation}

During training, we add loss terms (see Equation \ref{eq: PIM_loss}) to penalize the predicted $\alpha_1$ and $\alpha_2$ for the $i$th training point ($\hat{\alpha_{1,i}}$ and $\hat{\alpha_{2,i}}$, respectively) for deviating from their average values. The viscosity loss plus the penalty terms form the total loss $\mathcal{L}$.

\begin{equation}
    \mathcal{L} = \mathcal{L}_{\eta} + \frac{1}{n} \sum_{i=1}^n w_{\alpha}[(\hat{\alpha_{1,i}} - 1)^2 + (\hat{\alpha_{2,i}} - 3.4)^2]
\label{eq: PIM_loss}
\end{equation}

$w_{\alpha}$ is a hyperparameter that controls the impact that known values of the $\alpha_1$ and $\alpha_2$ parameters have on the final loss.

\subsection{Machine Learning Approaches}

The PENN and ANN models were implemented in PyTorch \cite{pytorch}. All models were trained on the same 9:1 (Train:Test) split. Before training, the features and $\eta$ were scaled to a range of (-1,1). The polymer fingerprint, PDI, and temperature were scaled with the Scikit-Learn MinMaxScaler \cite{scikit-learn} to a range of (-1,1). The $\dot{\gamma}$ was scaled by first adding a small value of $10^{-5}$, taking the $\log_{10}$, and then scaling to (-1,1). $M_w$ was scaled by taking the $\log_{10}$ value and then scaling to (-1,1). For the PENN, $\log M_w$ and $\log \dot{\gamma}$ use the same scaling bounds as $\eta$.

Within the training set, a 10-fold cross-validation (CV) was used to ensure that the models did not overfit the training set.  The ANN and PENN models also had separate models trained for each CV split. Hyperparameter optimization was performed using the Hyperband \cite{hyperband} optimization algorithm over each CV fold for both the ANN and the PENN models, with RayTune \cite{raytune} implementations, respectively. The ANN and PENN models, both containing 4 layers (including 2 hidden layers), involved optimization of the same hyperparameters: layer 1 size (64, 128, 256, 512), layer 1 dropout (0,0.01, 0.015,0.02,0.025,0.03), layer 2 size (64, 128, 256, 512), layer 2 dropout (0,0.01,0.015,0.02,0.025,0.03), and weight decay (0.00001, 0.00005, 0.0001, 0.0005, 0.001). For the PENN, $w_{\alpha}$ (0.001, 0.005, 0.01, 0.03, 0.05) was also optimized. The value corresponding to the lowest $\mathcal{L}_{\eta}$ (Equation \ref{eq: visc_loss}) of the CV test split was used. 

The Adam optimizer was used to train the models with a learning rate (LR) reduction by a factor or 0.5 on the plateau of the validation loss given a patience of 20 epochs. An initial LR of 0.0001 was used for the PENN. Empirically, we found that the PENN tuning was sensitive to high LR. The initial LR for the ANN was 0.001. Training was stopped with an Early Stopping patience of no improvement in the validation loss after 25 epochs.

The GPR model was implemented using Scikitlearn \cite{scikit-learn} trained using Bayesian optimization to tune key hyperparameters. The hyperparameters optimized include the noise level (\texttt{alpha}) with a range of $[10^{-2}, 10^{1}]$, the length scale of the RBF kernel (\texttt{length\_scale}) with a range of $[10^{-2}, 10^{2}]$, and the constant value used in the kernel (\texttt{constant\_value}) with a range of $[10^{-2}, 10^{2}]$, each with a logarithmic uniform prior. The optimization was performed over 50 iterations each over the 10-fold cross-validation, with the best-performing model parameters selected based on the results. The scaling for the inputs and outputs of the GPR were the same as the ANN.

\section{Data Availability}

The dataset used in this study is available at on the Ramprasad group's github (\url{https://github.com/Ramprasad-Group/polyVERSE}).

\section{Acknowledgements}

This work was financially supported by the Office of Naval Research (ONR) through Grant N00014-21-1-2258 and the National Science Foundation (NSF) DMREF Grant 2323695. A.J. acknowledges S. Shukla for help with the fingerprinting of polymer blends.

\section{Author Contributions}

 A.J. is the main architect of the models and dataset and wrote the paper. R.G. helped support the writing and technical soundness of this work. A.R. supported the data collection and analysis of the initial models. R.R. conceived the project and guided the work. 

\bibliography{achemso-demo}

\end{document}


\linenumbers
\newpage
\section{1. Additional Parity Plots for Test Set}

\begin{figure*}[!]
    \centering
    \includegraphics[scale=0.7]{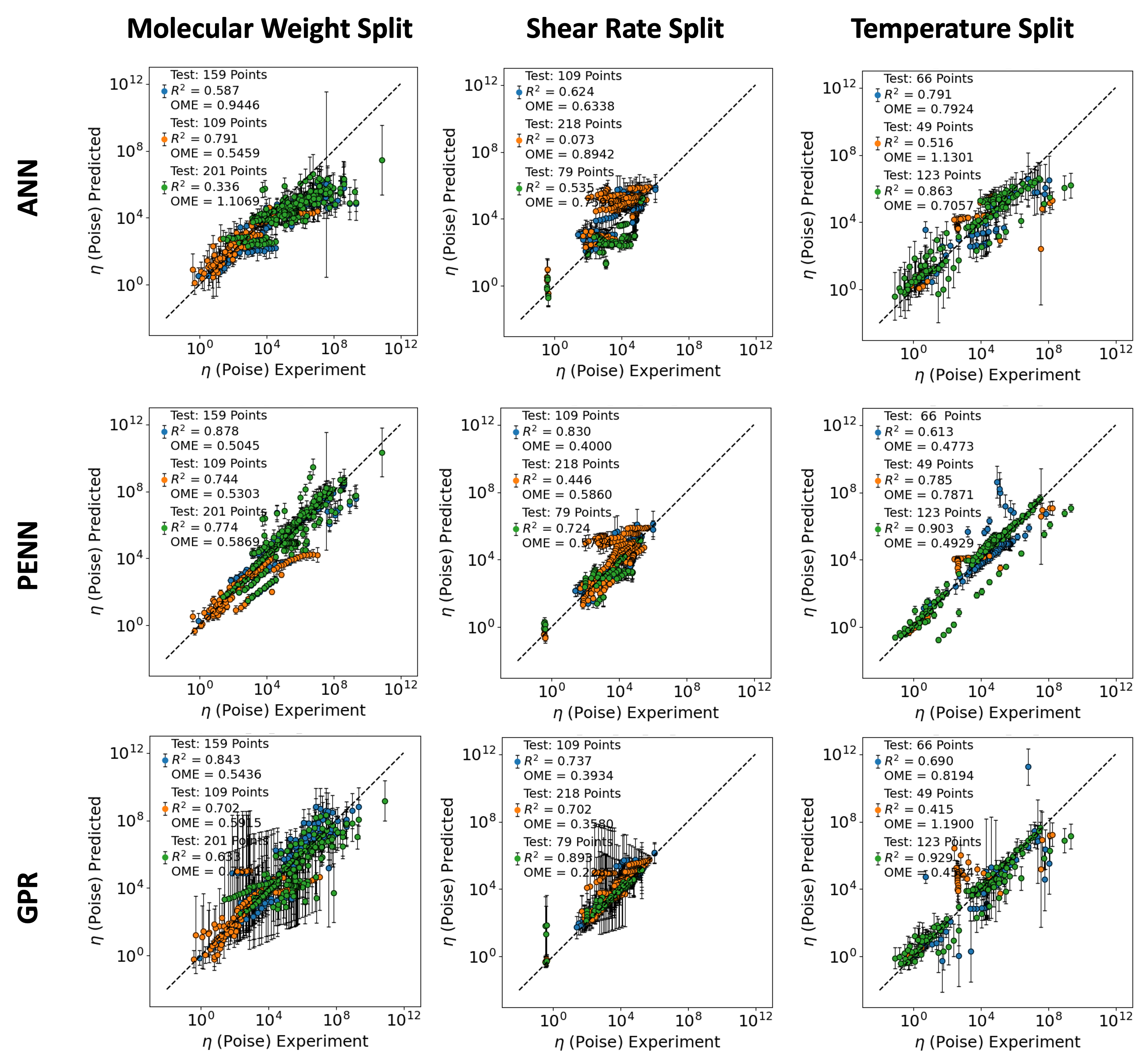}
    \caption{Parity plots containing trial information and the test sizes from each trial. Each plot compares experimental values for melt viscosity ($\eta$) to the predicted $\eta$. The dotted black lines represent perfect predictions. The coefficient of determination ($R^2$) and Order of Magnitude Error (OME) are reported over each test set trial.}
    \label{fig:parity_trials}
\end{figure*}

\clearpage


\section{2. Additional Extrapolation Plots in Various Test Cases}

\begin{figure*}[!htbp]
    \centering
    \includegraphics[scale=0.70]{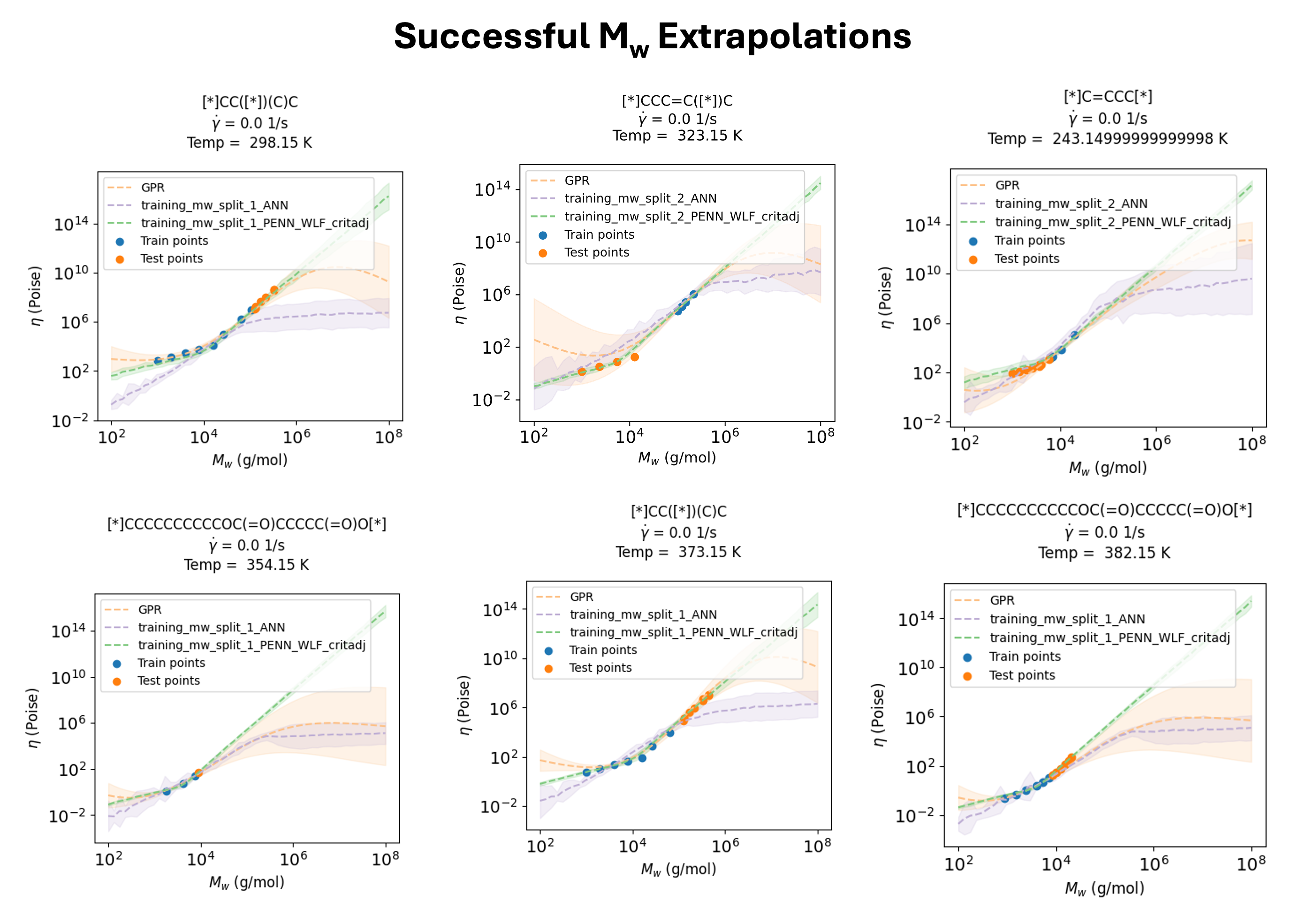}
    \caption{Examples successful of molecular weight extrapolations on partially seen and unseen monomers.}
    \label{fig:Mw_good_extrap_SI}
\end{figure*}

\begin{figure*}[!htbp]
    \centering
    \includegraphics[scale=0.70]{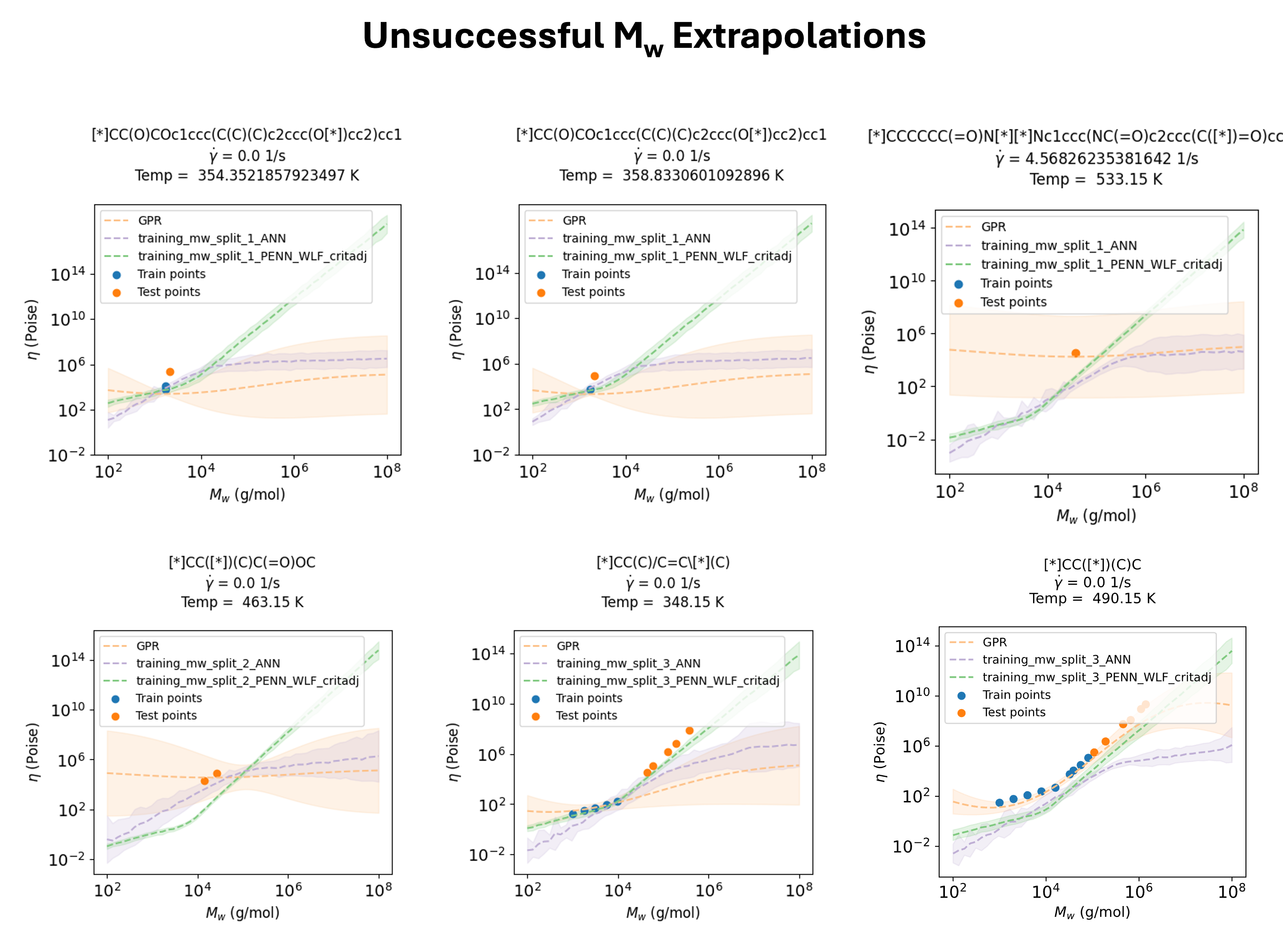}
    \caption{Examples of unsuccessful molecular weight extrapolations on partially seen and unseen monomers.}
    \label{fig:Mw_bad_extrap_SI}
\end{figure*}

\begin{figure*}[!htbp]
    \centering
    \includegraphics[scale=0.80]{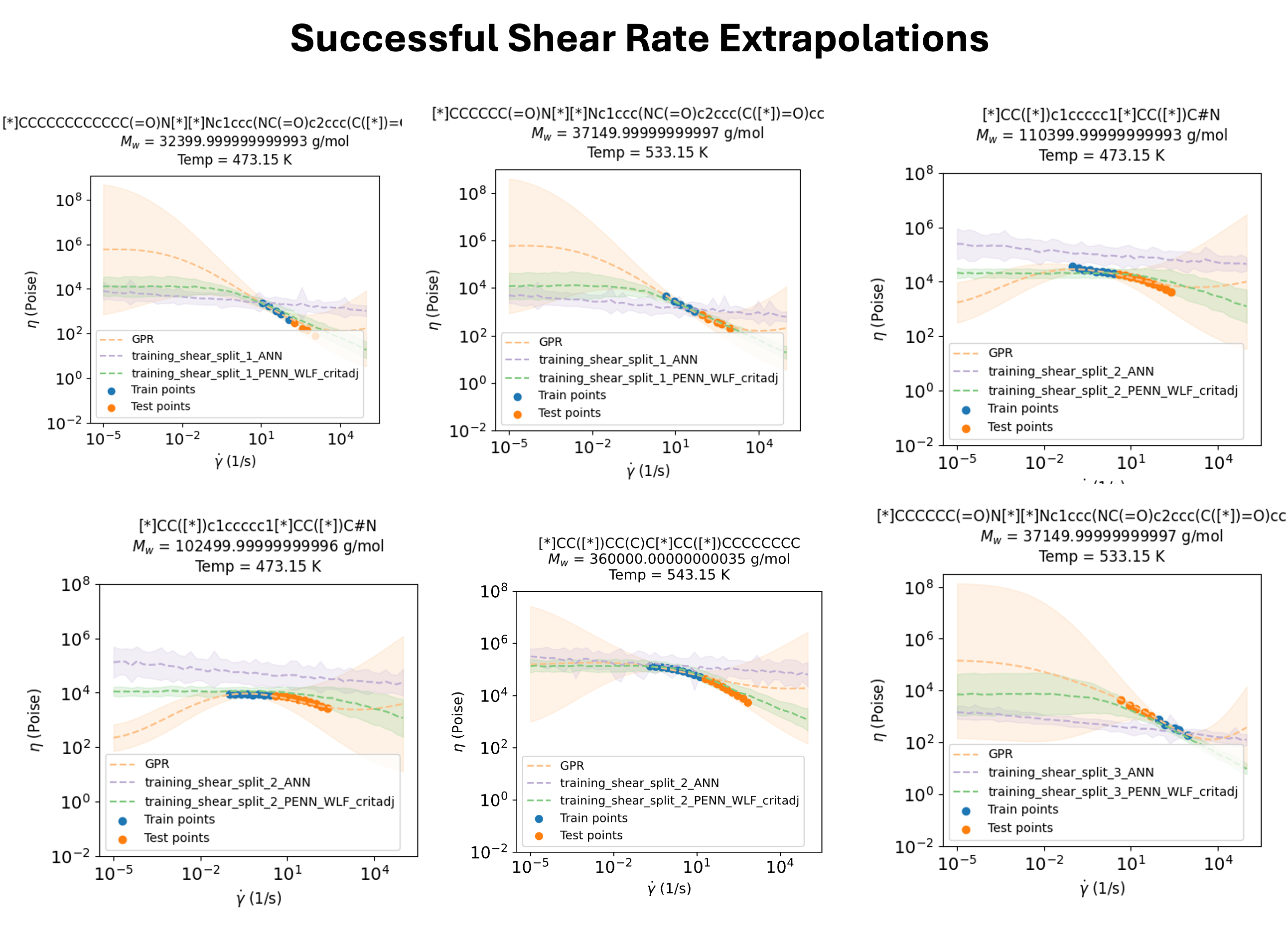}
    \caption{Examples of successful shear rate extrapolations on partially seen and unseen monomers.}
    \label{fig:Shear_good_extrap_SI}
\end{figure*}

\begin{figure*}[!htbp]
    \centering
    \includegraphics[scale=0.80]{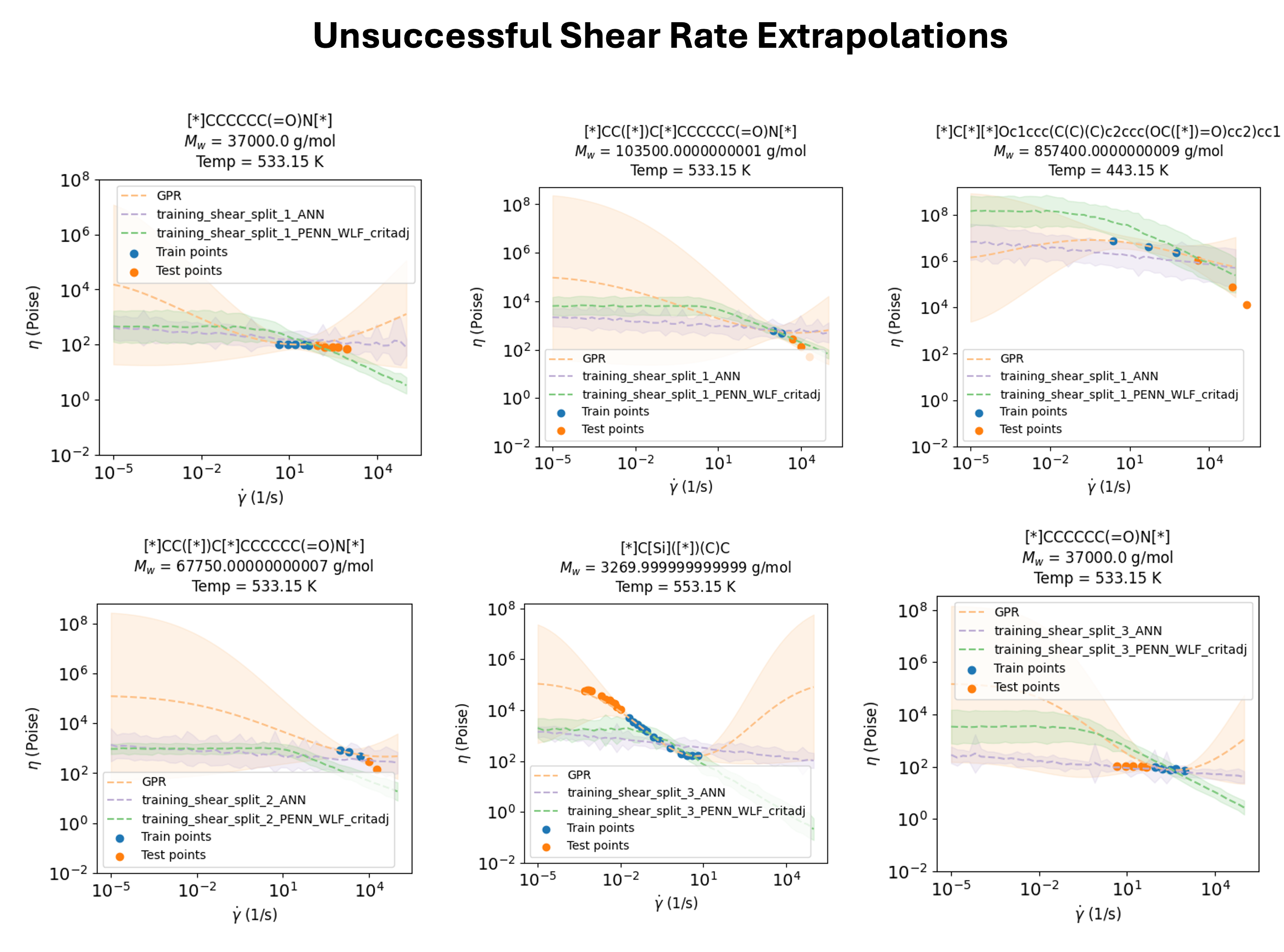}
    \caption{Examples of unsuccessful shear rate extrapolations on partially seen and unseen monomers.}
    \label{fig:Shear_bad_extrap_SI}
\end{figure*}

\begin{figure*}[!htbp]
    \centering
    \includegraphics[scale=0.80]{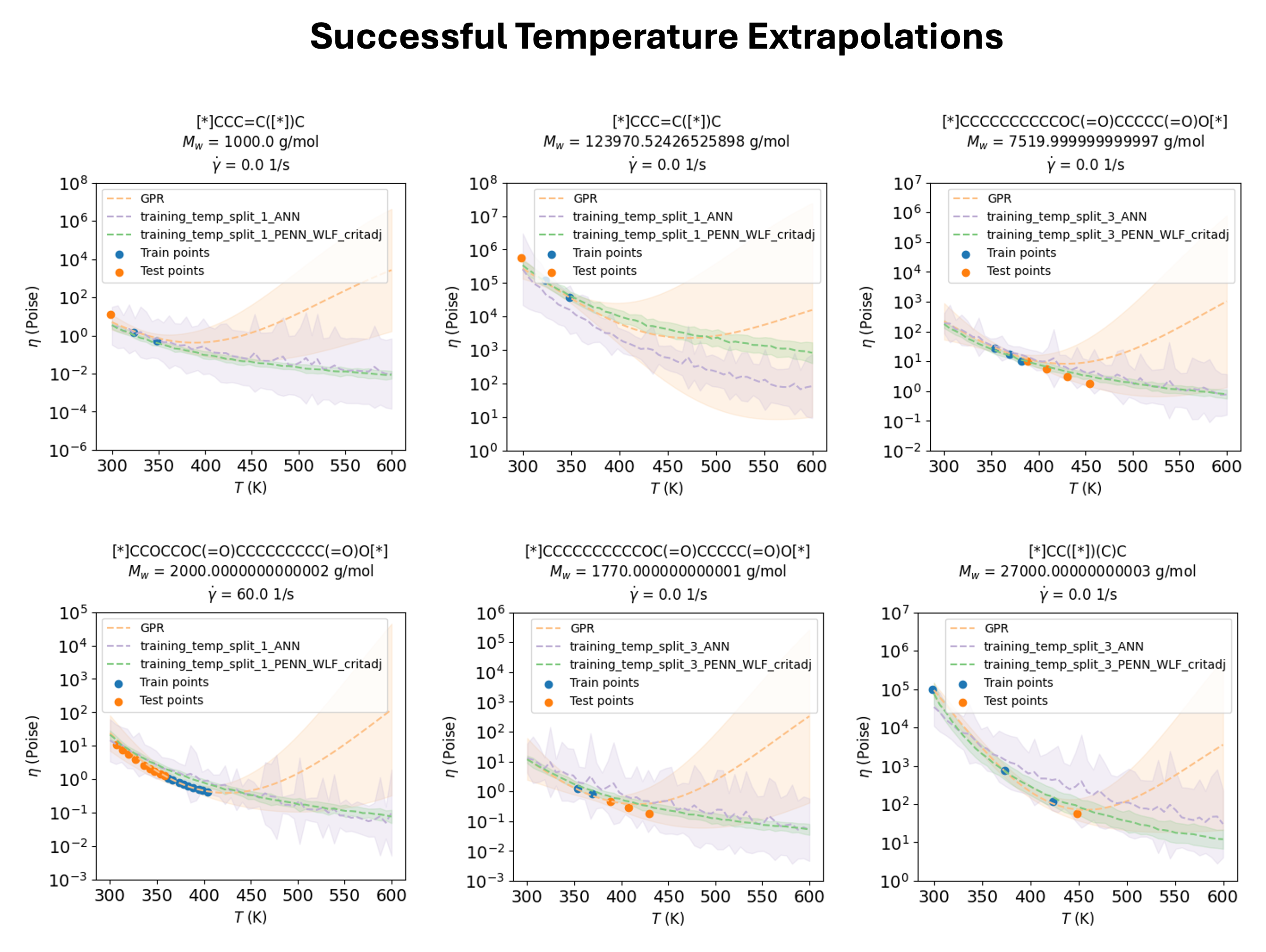}
    \caption{Examples of successful temperature extrapolations on partially seen and unseen monomers.}
    \label{fig:Temp_good_extrap_SI}
\end{figure*}

\begin{figure*}[!htbp]
    \centering
    \includegraphics[scale=0.80]{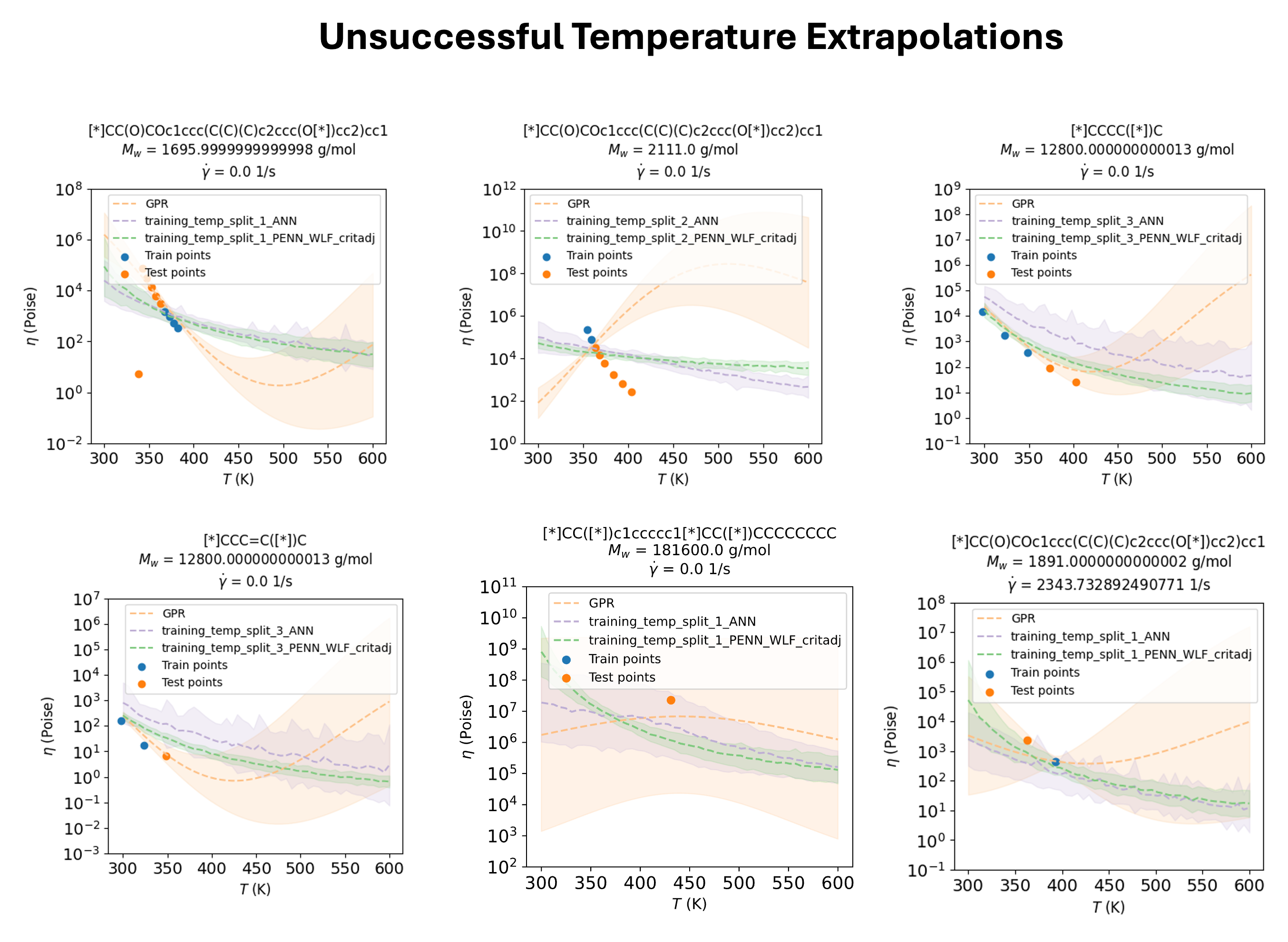}
    \caption{Examples of unsuccessful temperature extrapolations on partially seen and unseen monomers.}
    \label{fig:Temp_bad_extrap_SI}
\end{figure*}

\clearpage

\section{3. Bounding Ranges of Physical Parameters within the PENN Framework}

The final constants output from the MLP use the Sigmoid and Hyperbolic Tanget functions to hold them to physically meaningful ranges. The ranges also help reduce possible imbalances of gradients and/or exploding gradients that may occur during backpropogation, because of the complexity of the computation graph. In this work, we use a rudimentary approach to solve this problem, described in Table \ref{tbl:PIM_const_bounding} and more complete solutions may be developed in future works.

\begin{table}[h!]
  \caption{Bounded Ranges and Justifications of Empirical within the PENN Framework}
  \label{tbl:PIM_const_bounding}
  \begin{center}
  \begin{tabular}{p{20mm}p{25mm}p{90mm}}
    \hline
    Parameter & Bounding Range & Justification \\
    \hline
    $M_{cr}$   & (-1,1) & Keep critical value within $M_w$ ranges of the dataset \\
    $\alpha_1$  &  (0,3) & Bound to practical value near 1 \\
    $\alpha_2$  & (0,6) & Bound to practical value near 3.4\\
    $k_1$   & (-1.5,0.5) & Keep viscosity value within $\eta$ ranges of the dataset \\
    $\beta_M$  & (20, 50) & Appropriate range to control transition region, found through trial-and-error \\
    $C_1$ & (0,2) & Keep within practical ranges with regards to temperature scaling \\
    $C_2$ & (0,2) & Keep within practical ranges with regards to temperature scaling \\
    $T_r$ & (-1.5,1) & Keep reference temperature within $T$ and just below $T$ range of dataset\\
    $\dot{\gamma}_{cr}$  & (-1, 1) & Keep critical value within $\dot{\gamma}$ ranges of the dataset\\
    $n$ & (0,1) & Keep slope within range for general shear thinning fluids\\
    $\beta_{\dot{\gamma}}$  & (30) & Appropriate range to control transition region, found through trial-and-error\\
    \hline
  \end{tabular}
  \end{center}
\end{table}